\documentclass[reprint,superscriptaddress,11pt,onecolumn,nofootinbib,amsmath,amssymb,longbibliography[]{revtex4-1}

\usepackage{amsmath,color}
\usepackage{graphicx}
\usepackage{dcolumn}
\usepackage{bm}
\usepackage{amsfonts}
\usepackage{amsthm}
\usepackage{comment}
\usepackage{ulem} 
\usepackage{breqn}
\usepackage{rsfso}
\newcommand{\Dqparr}{D_{q_\parallel}}
\newcommand{\Sparr}{S_{q_\parallel}}

\usepackage[usenames,dvipsnames]{xcolor}
\usepackage[colorlinks,citecolor=BlueViolet,linkcolor=RedOrange,urlcolor=BlueViolet]{hyperref}

\begin{document}
	\title{Liquid-solid friction on crystalline surfaces}
	\author{Mathieu Lizée}
	\affiliation{Laboratoire de Physique de l'Ecole Normale Supérieure}
	\footnote{mathieu.lizee@ens.fr}
	\author{Alessandro Siria}
	\affiliation{Laboratoire de Physique de l'Ecole Normale Supérieure}

\begin{abstract}
Liquids flowing against solid surfaces experience friction. While solid friction is familiar to anyone with a sense of touch, liquid friction is much more exotic. At macroscopic scales indeed, the assumption of inifinite friction,  i.e. that interfacial liquid molecules stick to solid surfaces, is hard to disprove. Still, it has been known for a few decades that some materials exhibit very strong liquid slippage, leading to a dramatic increase in the permeability of nanoscale tubes to liquid flow. Harnessing liquid friction holds the promise of high-efficiency membrane separation processes, heat recovery systems or blue energy harvesting, making it a highly strategic field for reducing carbon emissions and addressing the climate emergency. In this chapter we review the history of liquid-solid friction measurements, mainly driven by the advent of new techniques and materials. We highlight the most established results and point out some directions that seem to us to be particularly dynamic and promising for the field.
\end{abstract}\maketitle

	\section{Introduction}
	
		Liquid-solid friction occupies a special place in the field of tribology. While solid friction is part of everyone's experience and has been ubiquitous in engineering for millennia, liquid-solid friction is a much more exotic phenomenon that has only been the subject of quantitative study for a few decades. In fact, the common assumption that liquids have zero velocity near solid surfaces holds surprisingly well for a wide range of materials, liquids and flow geometries, making infinite friction the standard and most common assumption. Deviations from this behaviour, in particular what is known as 'liquid slippage', have only been demonstrated in recent decades. Based on the development of very specific and diverse experimental platforms, the study of liquid-solid interface friction (or simply 'liquid friction') is still a young field of research, which is beginning to unveil an impressive wealth of effects, following in this way the broad phenomenology of solid-solid friction. \newline 
	
		We will begin by reviewing the main techniques for measuring liquid friction and discussing their respective contributions to the field. We will highlight important results such as the link between friction and wettability, the liquid superlubricity of graphene or the dependence of friction on temperature. We will then give some theoretical insights using liquid state theory and rationalise the link between slippage and contact angle. Finally, we will highlight several topics that seem to be particularly dynamic in the field and that the authors believe should be the focus of further work. In particular, we will discuss friction in ultra-confined environments, slippage on charged surfaces and the fate of liquid friction when the liquid is supercooled towards the glassy state.
	\newline
	
		Although the technological implications of liquid friction are beyond the scope of this chapter, they are crucial in the field of membrane processes. Indeed, slippery membrane materials could dramatically reduce energy costs and change the paradigm of separation processes, which are often based on highly energy-intensive distillation. For a review of membrane processes, see for example Ref.\cite{drioli2016encyclopedia}.
	
	\paragraph*{Slip length}
	To account for the possible mobility of liquid molecules near the surface and go beyond the infinite friction picture, Navier introduced the liquid-solid friction coefficient :
	\begin{equation}\label{constitutive_lambda}
		\lambda = \frac{\sigma_{interface}}{v_x}
	\end{equation}
	Eq.\ref{constitutive_lambda} is a Newtonian hypothesis for the liquid-solid interface in analogy to the bulk Newtonian assumption relating tangential stress to velocity gradients $\sigma = \eta \partial_z v_x$ where $\eta$ is independent of the tangential stress itself.
	
	Writing the continuity of tangential stress with the bulk liquid yields
	\begin{equation}\label{stress_continuity}
		\lambda =\frac{\sigma_{interface}}{v_x}= \frac{\eta \partial_z v_x}{v_x}
	\end{equation} 
	
	From Eq.\ref{stress_continuity}, we define the slip length as the distance behind the wall at which the velocity profile extrapolates to zero : $b = \eta/\lambda$ as sketched on Figure \ref{friction_techniques_resume} \textbf{a}. This length scale is key in estimating the impact of surface friction on fluid transport. Indeed, solving the Poiseuille problem for a cylinder of radius $R$ with this boundary condition leads to an increase of flux of order $b/R$. We will see in the following that slip lengths are nanometric for water on flat surfaces, meaning that liquid-solid friction controls transport in nanoscale channels but is largely negligible in wider geometry. This last observation explains why the study of liquid friction is tightly bound to liquid flow measurement in nanometric confinement, firstly in biological trans-membrane proteins and more recently in artificial channels.

	\section{Experimental techniques}
\begin{figure*}
		\centering
		\includegraphics[width=\linewidth]{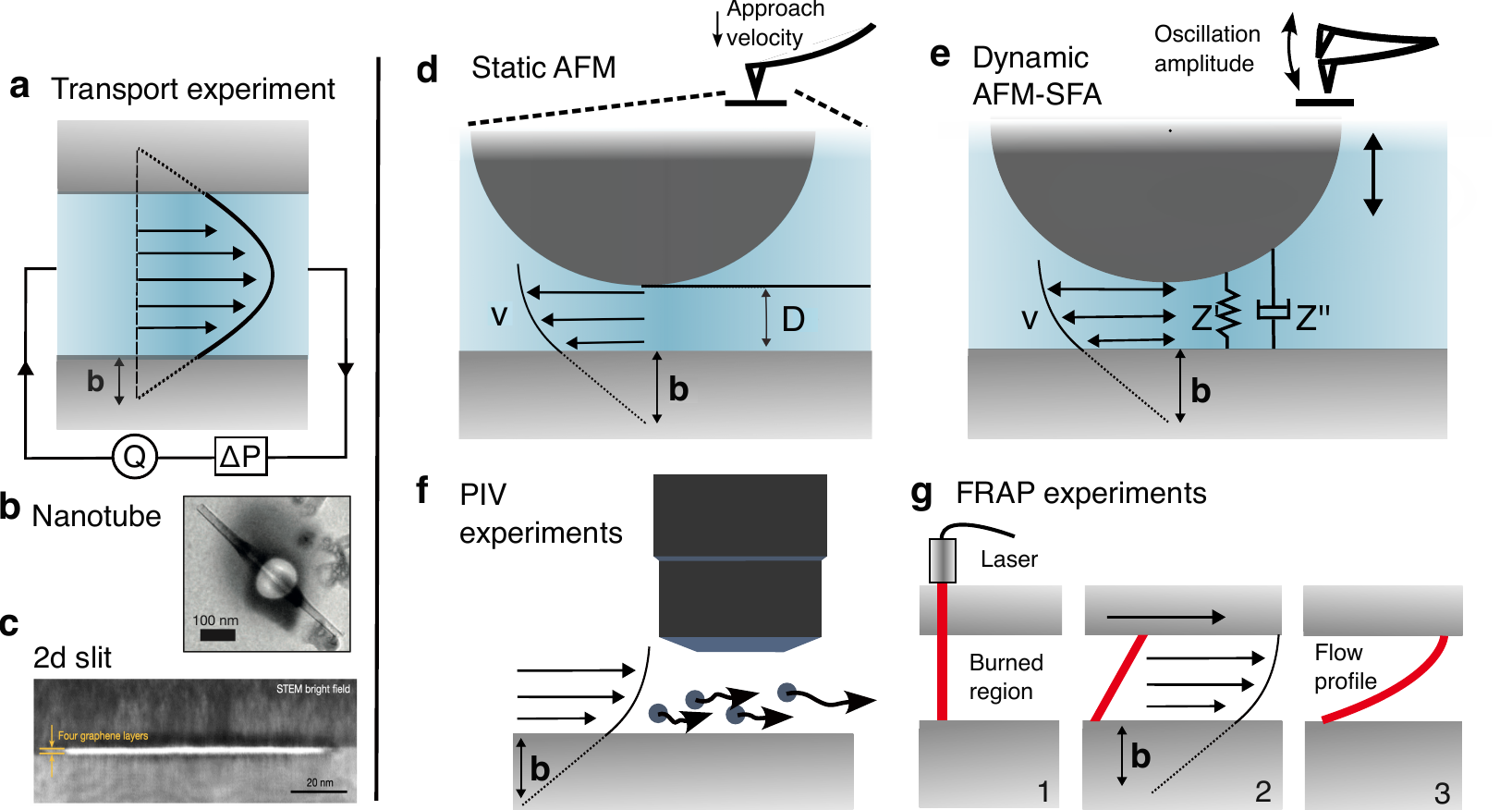} 
		\caption{\textbf{Overview of the main experimental techniques used to measure liquid-solid friction} On panel \textbf{a}, we define the slip length $b$, in the context of pressure driven water transport, as the distance inside the solid at which the linear extrapolation of tangential velocity vanishes. One should note that whereas positive slip length denotes a non zero tangential velocity at the solid surface, negative slip length makes sense as well as it occurs if a layer of liquid rests immobile at the surface. On panels \textbf{b-c}, we show electron microscope images of solid state crystalline nanochannels in which liquid friction can be extracted from transport measurements, respectively a transmembrane nanotube on \textbf{b} \cite{siria2013giant} and a 2d nanometric slit \cite{radha2016molecular} on \textbf{c}. On panels \textbf{d-e}, we describe liquid friction measurements with static and dynamic atomic force microscopes (AFM). In the first case, the sphere approaches the surface at constant velocity. In the dynamic mode however, a nanometric oscillation allows to extract the complex mechanical impedance in confinement at the oscillation frequency. In each case, an approach curve is used to infer slippage. On panel \textbf{f}, we sketch the PIV experiments: tracer particles allow to image the liquid flow close to the wall. The FRAP technique is similar but requires no tracers (\textbf{g}), a laser marks the surface and thus materializes the velocity profile. One should be aware that there are various different FRAP schemes and that we sketch here the one which has produced most of recent results.}
		\label{friction_techniques_resume}
	\end{figure*}

	In this section, we will briefly sketch the history of liquid friction measurements, and introduce the dedicated techniques and instruments along with their respective strengths and weaknesses. On Figure \ref{friction_techniques_resume} \textbf{a-c}, we show the principle of slippage measurement from transport experiment and electronic microscope images of BN-nanotubes and 2d nanochannels from Refs.\cite{siria2013giant,yang2020capillary}. On \textbf{d}, we sketch the static CP-AFM technique in which a sphere glued at the end of a cantilever approaches a surface at fixed velocity while the lever's deflection is recorded. The alternative dynamic AFM/SFA scheme, sketched on \textbf{e}, takes advantage of a vertical oscillation to extract the dissipative impedance caused by viscous dissipation in an axissymmetric drainage flow. On panels \textbf{f-g}, we show the concepts of particle imaging velocimetry (PIV) and Fluorescence Recovery After Photobleaching (FRAP) techniques which image the liquid flow profile to infer slippage. Note that many variants of the FRAP technique have been implemented. The one described here corresponds to the setup of Restagno \textit{et al.} used in Ref.\cite{henot_temperature-controlled_2018}.

	\subsection{Transport measurement}\label{subsec:transport}
	Transport enhancement (\textit{cf.} Figure \ref{friction_techniques_resume} \textbf{a-c}) is a very natural way to measure slippage and was therefore one of the first techniques successfully used for this purpose \cite{churaev1984slippage}. For example, the mass flux through a cylindrical channel strongly depends on slip as follows : 
	\begin{equation}
		Q_c = \frac{\rho \pi R^4}{8 \eta L}\Bigl( 1+\frac{4b}{R}\Bigr) \times \Delta P
	\end{equation} where $\eta$ is the liquid viscosity, $R$ and $L$ the channel's radius and length. To ensure a measurable slip-induced flow enhancement $4b/R$, one needs $R \sim b$ which is typically nanometric for water. To realize such nanometric confinement one can fabricate a nanoporous material and measure its permeability averaged over thousands of channels. An estimation of porosity then allows to infer liquid-solid friction. In the following, we will use the seminal example of carbon nanotubes (CNT). Permeability measurements on CNT membranes have brought the first hints of ultra-low water friction \cite{majumder2005enhanced,holt2006fast} and attracted a considerable interest in the field. Experiments in such membranes however raise legitimate questions: How to evaluate the active surface area or avoid leaks ? Is there any cross-talk between these closely packed tubes or are there independent? Moreover, contamination by a linking agent or the catalyst used for nanotube growth has been reported to hinder measurements. The considerable scatter of CNT membrane permeability has required single-channel experiments which are the object of the next discussion.
	
	
	\begin{figure*}
		\centering
		\includegraphics[width=\linewidth]{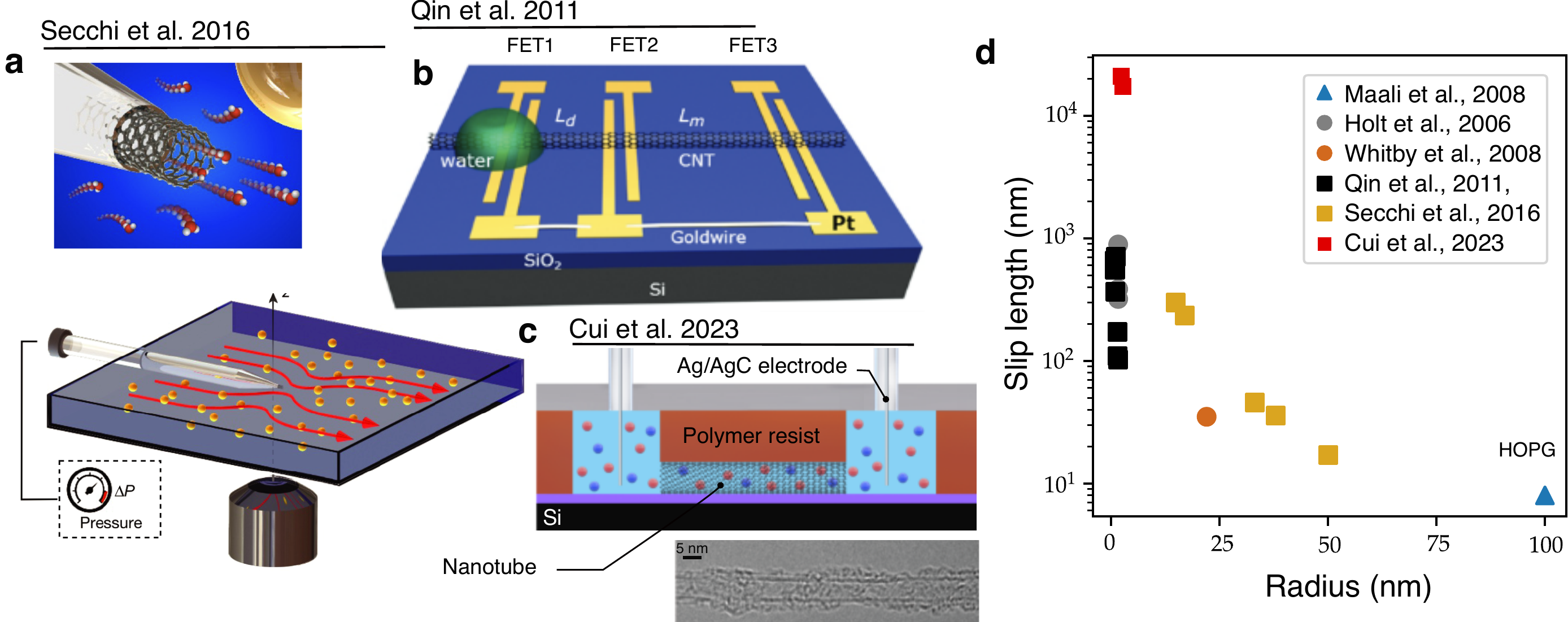} 
		\caption{\textbf{Slippage in carbon nanotubes} On panel \textbf{a}, we show the experimental setup used by Secchi \textit{et al.} in Ref.\cite{secchi_massive_2016}. The tube is inserted in a glass capillary pipette. As a pressure is applied in the pipette, water flow is imaged by PIV by imaging micrometric colloids under a microscope. On panel \textbf{b}, reproduced from \cite{qin2011measurement}, we sketch the experimental system used by Qin \textit{et al}. A droplet is placed on an end of a nanotube crossing several gold contacts. After cracking the tube open with FET1, wetting velocity in the tube is probed between FET2 and FET3 and the interfacial friction coefficient is subsequently inferred. On panel \textbf{c}, we sketch the nanotube system made in the group of Ming Ma \cite{cui2023enhanced}. Voltage and concentration drops are applied across the nanotube while the ionic current is recorded between the electrodes. A TEM image of the double wall tube is also shown. Finally, we summarize on panel \textbf{d} results from the best controlled experiments. A drastic increase of slippage upon decreasing diameter is clearly observed up to extremely high values of 20 $\mu$m from Ref \cite{cui2023enhanced}. Note that in most studies, the electronic properties of the tubes (semi-conducting or metallic character) are unknown.}
		\label{figure_CNT}
	\end{figure*}

	\paragraph*{Carbon nanotube friction}   
	Following the reports of ultra low friction in carbon nanotube membranes \cite{majumder2005enhanced,holt2006fast,majumder_mass_2011}, huge efforts were devoted to the study of water confined in carbon nanotubes and especially to the friction of water on the inner tube's wall.  We summarize the most important results on Figure \ref{figure_CNT} \textbf{d}, focusing in particular on the better controlled experiments : those on single nanotubes. For the sake of comparison, the results of Holt \textit{et al.} \cite{holt2006fast} on CNT membranes ($b\sim 1$ $\mu$m) is shown as a grey dot on panel \textbf{d}.
	
	\paragraph*{Tube filling dynamics}
	The single-tube permeability was first measured for very thin ($R < 1$ nm) nanotubes deposited on a silicon wafer by Qin \textit{et al.} \cite{qin2011measurement} (\textit{cf.} Figure \ref{figure_CNT} \textbf{b}). The tube can be opened by a high current between the left electrodes (FET 1). The water-front is then detected by conductivity measurement at several positions (FET 2 and 3), allowing to estimate the filling velocity. Unfortunately, as the driving force was not precisely known, the flow profile remains somehow uncertain. Nevertheless, the obtained slip lengths ranging from 100 nm to 1 $\mu$m are shown as black squares on our summary panel \textbf{d}.
	
	\paragraph*{True flow measurement with PIV detection}
	A few years later, Secchi \textit{et al.}  inserted tubes in a quartz nanopipette. This geometry allowed to track the pressure-driven mass flow with microscopic colloids close to the tube's opening \cite{secchi_massive_2016} (\textit{cf.} Figure \ref{figure_CNT} \textbf{a}). The authors reported a very low and diameter-independent slippage on boron-nitride nanotubes. Interestingly, the results were radically different for carbon : not only is slippage extremely high but it is also strongly diameter dependent (yellow squares on panel \textbf{d}). These results are consistent with the massive permeability of 2 nm carbon nanotube membranes \cite{majumder2005enhanced,holt2006fast}. The strong discrepancy between carbon and boron-nitride despite identical crystallographic properties is believed to be caused by a difference in the mobility of adsorbed OH$^-$ ions \cite{mangaud2022chemisorbed}. The dramatic increase of slippage in smaller tubes is no less puzzling and it was recently proposed that differences in the electronic structure of walls of different radii could be the cause of this discrepancy \cite{kavokine_fluctuation-induced_2022} (see subsection \ref{subsec:carbon_friction} for details). 
	
	\paragraph*{Enhanced control on the tubes properties}
	In a very recent work, the authors managed to build a microfluidic system to address a single double-walled nanotube \cite{cui2023enhanced} (\textit{cf.} panel \textbf{c}) with a 5 nm inner diameter and well established electronic properties. Although the geometry does not allow to apply large pressure drops across the channel, one can apply concentration and voltage drops across the tube and measure the induced ionic current. Upon several assumptions, the authors estimated the slip length to be as high as 20 $\mu$m in semi-conducting tubes, corresponding to a dramatic drop in friction. These measurements correspond to the red squares on panel \textbf{d}.
	
	
	
	\paragraph*{Intrinsic weakness of transport measurements}
	The main difficulty with transport measurements is their non-local nature that makes them sensitive to any impurity, contamination, or surface heterogeneity. They also require very elaborate and often material-specific fabrication techniques. Moreover, even once the nanofluidic channel has been fabricated, friction sensing still requires an exquisitely sensitive mass flow measurement under a pressure gradient. The great difficulty and intrinsic weaknesses of this experimental approach therefore call for the adoption of a more versatile scheme, where friction can be accessed for any liquid/solid pair.

	

	\subsection{Hydrodynamic flow imaging : FRAP and PIV}
	Among the early attempts to a more versatile liquid friction measurement scheme, the Fluorescence Recovery After Photobleaching (FRAP) was developed in the group of Liliane Léger \cite{pit2000direct}. This technique, which can be viewed as a tracer-less Particle Imaging Velocimetry (PIV) scheme, offers a local measurement and provides the full flow profile in a planar geometry. A few years later, promising PIV-based slippage measurements have yielded a slip length of water on silanized glass of $b\sim 16$ nm \cite{joly2004hydrodynamics}. The FRAP scheme has further been used in recent years with a new configuration, detailed on Figure \ref{friction_techniques_resume} \textbf{g}, to measure the friction of polymer melts as a function of temperature and chain length \cite{henot_temperature-controlled_2018}. A laser is used to burn a line in a fluorescent polymer (functionalized PDMS). A shear motion is then applied, allowing to image the flow and infer slippage. In this study, Henot \textit{et al.} tried to disentangle the two contributions to the slip length $b=\eta/\lambda$ : viscosity and liquid-solid friction. By varying temperature, the authors identified two independent activation energies for viscosity ($E^{\eta}_a$) and liquid-friction coefficient ($E^{\lambda}_a$). They report the independency of $E^{\lambda}_a$ on chain-length $N$ and conclude that liquid friction is controlled by the monomer's structure only, in drastic contrast with viscosity which depends on chain length as $\eta \sim N^3$. An interesting consequence, already pointed out by de Gennes \cite{de2003ecoulements}, is that slip length of polymer melts is proportional to viscosity while it is expected to be independent of $\eta$ in simple liquids \cite{bocquet2010nanofluidics}. Overall, the PIV and FRAP techniques have several drawbacks : they require a functionalization step or the addition of tracers to liquids and their resolution is much lower than dynamic SFA and colloidal probe-AFM. However, the simplicity of the FRAP process is a major great advantage that has allowed Restagno \textit{et al.} to collect extensive data on polymer friction.
	
	
	\subsection{Surface Force Apparatus}
	
	The investigation of the liquid-solid boundary condition has taken a new turn in the 1970's with the introduction of the Surface Force Apparatus (SFA). The SFA has allowed to routinely confine various liquids between atomically smooth surfaces while measuring shear or normal forces \cite{israelachvili2010recent}. Beyond the measurements of surface forces, the SFA was also used to probe confined hydrodynamics and in particular interfacial slippage. Early experiments \cite{chan1985drainage,georges1993drainage} have set a landmark in confined liquids physics by showing that continuous hydrodynamics hold down to the molecular scale (1 nm) which is in itself a most surprising result. In these works, the authors report an immobile layer of molecules adsorbed at the surface resulting in a negative slip length.
	
\begin{figure*}[h!]	\centering
		\includegraphics[width=\linewidth]{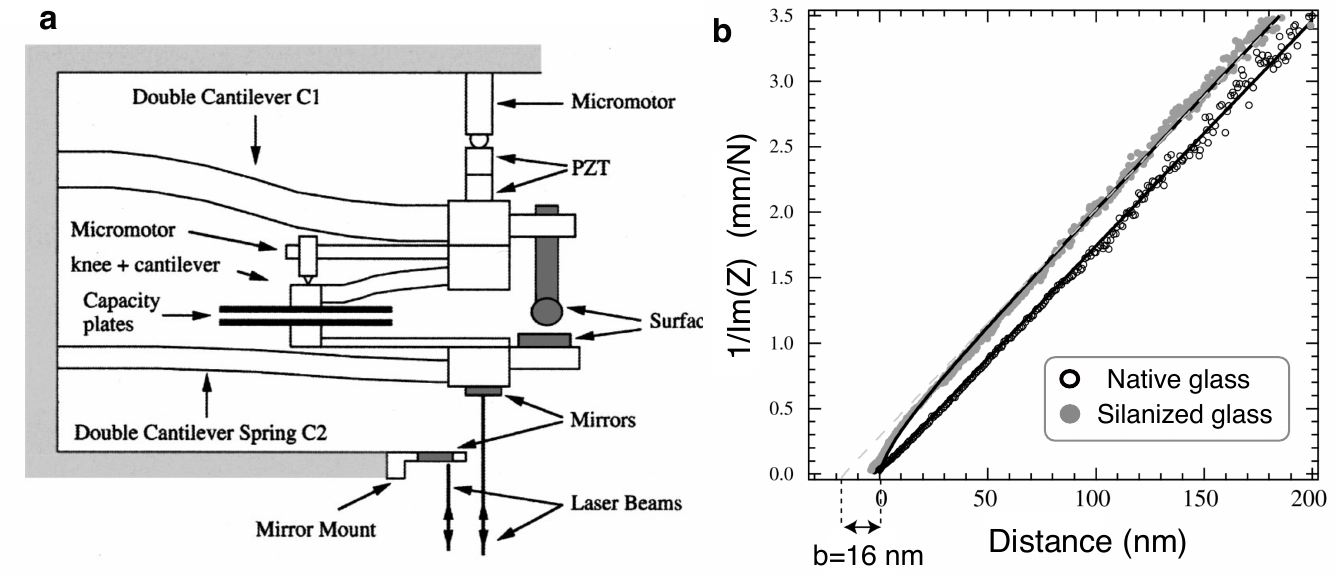} 
		\caption{\textbf{Dynamic Surface Apparatus (dSFA)} used to probe nano-confined liquids rheology and interfacial friction. Panel \textbf{a} decribes the system and is extracted from \cite{restagno2002new}. On panel \textbf{b}, we reproduce the curves of the inverse dissipation \textit{versus} distance on native and on silanized glass, extracted from \cite{cottin-bizonne_boundary_2005}. The curvature shown on silanized glass denotes a reduced friction at long distance and an increased slope of dissipation close to the contact for slipping liquids. This behavior is in excellent agreement with the theory by Vinogradova \cite{vinogradova_drainage_1995} and is a key element to discard the presence of possible artefacts -- like impurities -- wich could lead to errors in the slippage estimate.}
		\label{Figure_SFA}
	\end{figure*}

	\paragraph*{The dynamic SFA}
	To enable systematic slippage measurements with the SFA, a stiffer system was needed. Built around a glass sphere oscillating at a frequency $f\sim 10-300$ Hz, the dynamic SFA (d-SFA) built in the group of Elisabeth Charlaix has allowed to measure liquid friction for various kinds of surfaces. The dual detection mechanism (optical and capacitive) ensures a sub-nanometric resolution in both static and dynamical modes. The d-SFA is described elsewhere \cite{restagno2002new} and we only briefly sketch its design on Figure \ref{Figure_SFA} \textbf{a} and the measurement concept on Figure \ref{friction_techniques_resume} \textbf{e}. The millimetric sphere is brought in close proximity to a planar surface and probes a region of size $\sqrt{RD} \sim$ 10 $ \mu$m. The sphere is subjected to a vertical oscillation at a chosen frequency $\omega$ and nanometric amplitude $a$ while its distance to the plane is measured by interferometry. The viscous dissipation in the drainage flow between the sphere and planar surface is given by the Reynolds formula :
	\begin{equation}\label{eq:reynolds_chap_intro}
		\textrm{Im}(Z)=\frac{6 \pi \eta R^2 \omega}{D+b}
	\end{equation}
	where we assumed a no-slip boundary condition on the sphere and a slip length $b$ on the planar surface. The simplicity of this geometry allows to take into account a wide variety of effects in the force curves. For instance, the impact of slippage on the flow (correction to Eq.\ref{eq:reynolds_chap_intro}) \cite{vinogradova_drainage_1995}, bulk visco-elasticity \cite{cross_wall_2018} and elastic surface deformation \cite{leroy_hydrodynamic_2011} have been accounted for in an extended analytical theory, brilliantly confirmed by experiments.
	
	\paragraph*{Slippage measurements on silanized glass}
	The d-SFA can be credited of some of the most systematic slippage measurements, on various surfaces and different liquids, backed up by the thorough understanding of dissipation curves which depart from the Reynolds formula at short distance (\textit{cf.} Figure \ref{Figure_SFA} \textbf{b}). A major result is the quantitative confirmation that hydrophobicity promotes slippage, first observed by Churaev \cite{churaev1984slippage}, but measured unambiguously with the d-SFA by Cottin \textit{et al.} \cite{cottin-bizonne_boundary_2005}. In this experiment, the glass sphere faces either a bare hydrophilic glass plane or its silanized, strongly hydrophobic, counterpart. As shown on Figure \ref{Figure_SFA} \textbf{b}, the slip length for water is measured at zero for the native surface and 16 nm for the silanized one. Here, not only do the authors establish hydrophobic enhanced slippage, they also confirm with excellent resolution the no-slip boundary condition on an hydrophilic substrate of sub-nanometric roughness. Despite the excellent sensitivity of the d-SFA, it remains very hard to operate for two main reasons : (1) the long time stability required by the ultra-slow approach velocity and of (2) the large contact area making it very vulnerable to contamination. Another experimental technique for slippage measurement shows none of these shortcomings and is now established as the central technique for liquid-solid friction : the colloidal-probe atomic force microscope (CP-AFM).

	\subsection{Colloidal probe AFM}
	
	\begin{figure*}[h!]
		\centering
		\includegraphics[width=\linewidth]{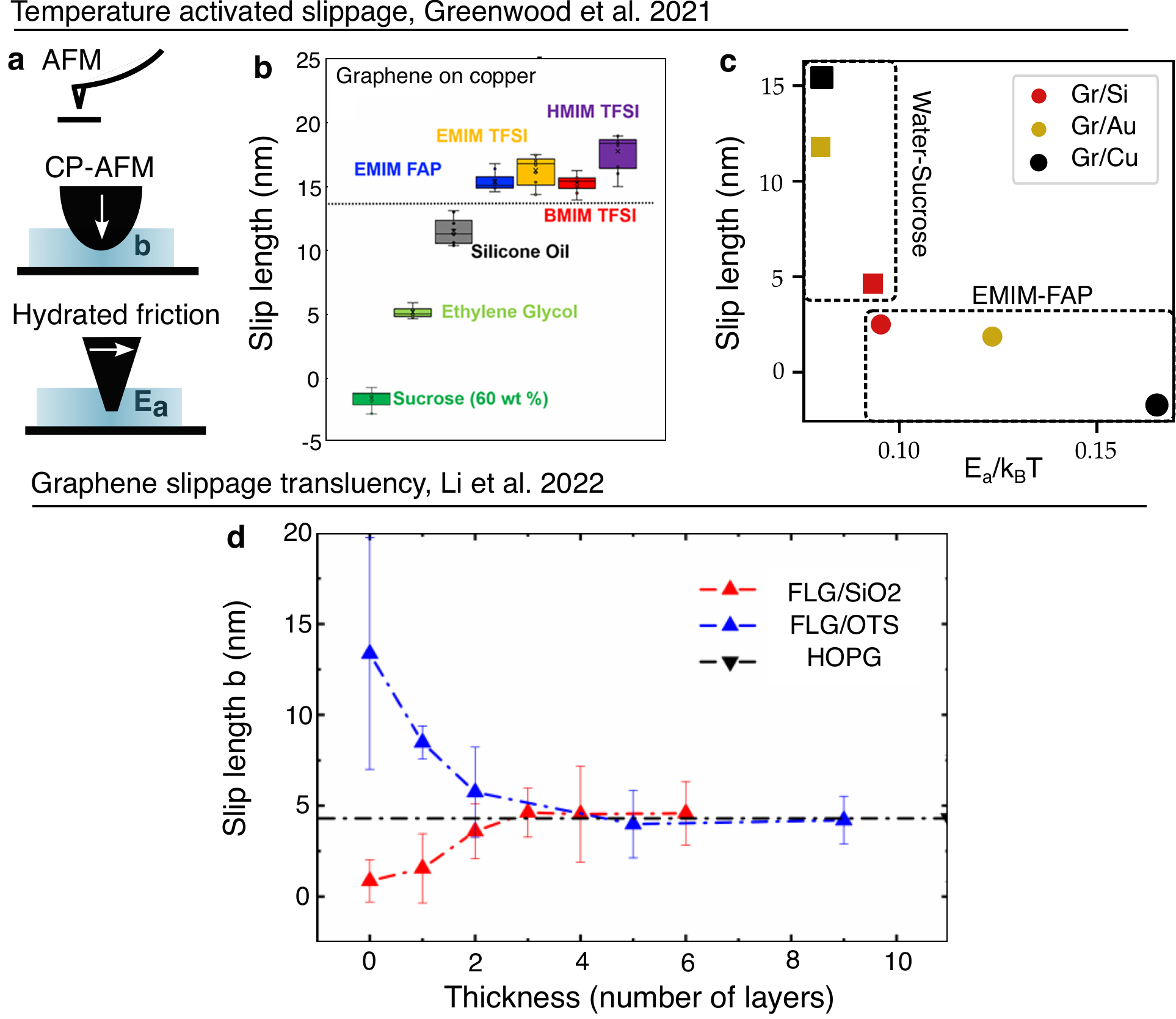} 
		\caption{\textbf{Recent experimental results obtained with the CP-AFM} On panel \textbf{a-c}, we plot results from the group of Espinosa-Marzal, \cite{greenwood_effects_2021}. In this work, the authors measure the slip lengh of various liquids on three different graphene surfaces. The originality of their work lies in the combination of liquid friction (CP-AFM) with hydrated solid friction.measurements under well defined tangential stress. These two experiments allow to plot on panel \textbf{c}, the negative correlation of slip length and local activation energy in the sense of Eyring. On panel \textbf{d}, we display the experimental results from the group of Ming Ma \cite{li_translucency_2022} on graphene crystals of various thickness. Here, graphene is deposited either on a hydrophillic ($\rm SiO_2$) or a hydrophobic substrate (OTS). The results show that graphene plays the role of a screening layer protecting the liquid flow from the underlying substrate while the latter still controls most of the liquid friction behavior.}
		\label{Figure_AFM}
	\end{figure*}
	
	One of the most prolific instrument in the measurement of liquid slippage is the atomic force microscope (AFM) and its micrometric sphere version, the colloidal probe AFM. A micrometric sphere is glued on an AFM cantilever and immersed in liquid close to the chosen surface. Whereas the SFA requires millimeter scale samples, the CP-AFM allows to probe friction on any material, including micrometric single crystals. The small probe size further ensures a good robustness against contamination and one can easily move on the sample surface to check reproducibility at various positions. Once the sphere is glued, the experimentalist approaches the flat surface at constant velocity $v\sim$ 10 - 100 $\mu$m/s. The friction force is given by the Reynolds equation $F \sim \eta R^2 v/(D+b)$ and allows to extract the slip length $b$ in complete analogy with the SFA.  For these reasons, the colloidal probe AFM has been extensively used to probe the dependency of slippage on surface properties like roughness, wettability or surface charges. In analogy to the dynamic SFA, the CP-AFM has also been used in dynamical mode \cite{bhushan2009boundary} and confirmed that $b\sim0$ for mica. The authors obtained $b= 43\pm10$ nm for an alcane self-assembled layer and report ultra high slippage on super-hydrophobic composite surfaces which are out of the scope of this chapter. In the following we focus on recent efforts to measure water friction on graphene with the CP-AFM.
	
	\paragraph*{From graphene to graphite friction}
	Among the important measurements which are hard to conceive without the CP-AFM is that of the 8 nm slip length of water on graphite by Maali \textit{et al.}.  Investigations of water-carbon friction have recently been pushed forward to few-layer graphene samples on $\rm SiO_2$ in a genuine experimental \textit{tour de force} \cite{li_translucency_2022}. In this study, the authors provide a measurement of water slippage as a function of graphene thickness -- down to the single layer resolution -- with hydrophilic SiO$_2$ and hydrophobic silanized surfaces. Their results point towards a transluency of graphene for liquid-solid friction and are reproduced on Figure \ref{Figure_AFM} \textbf{d}. For thinner flakes, the substrate dictates liquid friction and one obtains high slippage for silanized substrate while the no-slip boundary condition holds for $\rm SiO_2$. As the flake thickens, graphene increasingly screens the substrate and ultimately, graphite friction coefficient is recovered above a few nanometers in rough agreement with the 8 nm value of Maali \textit{et al.}.
	
	\paragraph*{Link with hydrated solid friction}
	In recent years, attempts to use the AFM to unify hydrated-solid friction and slippage measurements have brought extremely interesting results to the field. The authors of Ref.\cite{greenwood_effects_2021} apply Eyring's theory of transition states to liquid-solid friction \cite{eyring1936viscosity}. In this framework, interfacial diffusion is controlled by an activation energy $E_a$ and assisted by tangential stress. The activation energy can be measured using velocity-dependent hydrated friction : the solid friction between a sharp AFM tip and the surface is measured in the presence of a nanometric water lubrication layer as sketched on Figure \ref{Figure_AFM} \textbf{a}. Here, the tip is shearing the surface, exactly like in solid-solid friction AFM measurements. By combining these results with standard CP-AFM drainage experiment yielding slip length $b$ for various liquids on graphene (\textit{cf.} Figure \ref{Figure_AFM} \textbf{b}), the authors correlate liquid-solid friction with the activation energy $E_a$ (see panel \textbf{c}). Here, the authors investigate an aqueous sucrose solution and an ionic liquid (EMIM-FAP) on CVD graphene deposited on three different substrates : copper, gold or silicon. They report a drastic drop of slip length from 16 nm down to negative values as $E_a$ increased from 0.05 to 0.15 $k_{\rm B}T$. In another work, Diao \textit{et al.} have systematically investigated $E_a$ and $b$ for salted water solutions on graphene \cite{diao2019slippery}. This original approach to liquid-solid friction seems very promising to us as it disentangles the activated equilibrium contribution $E_a/k_{\rm B}T$ from the stress-assisted one which is often negligible in conventional drainage setups.

	\section{From numerical techniques to theoretical insights}
	From a theoretical perspective, the liquid-solid interface is a notoriously difficult problem, where statistical mechanics of liquids and solid-state physics mix together at surfaces.

	\subsection{Static wall friction}
		A first attempt to estimate liquid friction on a corrugated potential can be made within the linear response theory. Using the Green-Kubo formula, one relates the interfacial momentum transport coefficient to the auto-correlation function of the force (\textit{ie:} momentum flux) at equilibrium \cite{barrat1999influence} on a surface $\mathcal{A}$.
	\begin{equation}
		\lambda=\frac{1}{\mathcal{A}k_{\rm B}T}\int_{0}^{\infty}dt \langle f_x(t) f_x(0)\rangle
	\end{equation}
	
	The interfacial friction coefficient can eventually be rewritten in terms of the fluid-solid interaction potential, the liquid's structure factor at the interface $\Sparr$, its density $\rho$ and its typical lateral diffusion coefficient at the scale of corrugations $\Dqparr$, which can all be obtained from equilibrium molecular dynamics simulations.
	
	\begin{equation}\label{eqn:Green_Kubo}
		\lambda=\frac{\Sparr}{\Dqparr k_{\rm B}T}\int_{0}^{\infty}dz \rho(z) V_{FS}(z)^2
	\end{equation}
	
	The obtained slippage can then be compared with the velocity profile from out of equilibrium simulations \cite{thompson1990shear,barrat1999large}.
		\begin{figure*}[h!]
		\centering
		\includegraphics[width=\linewidth]{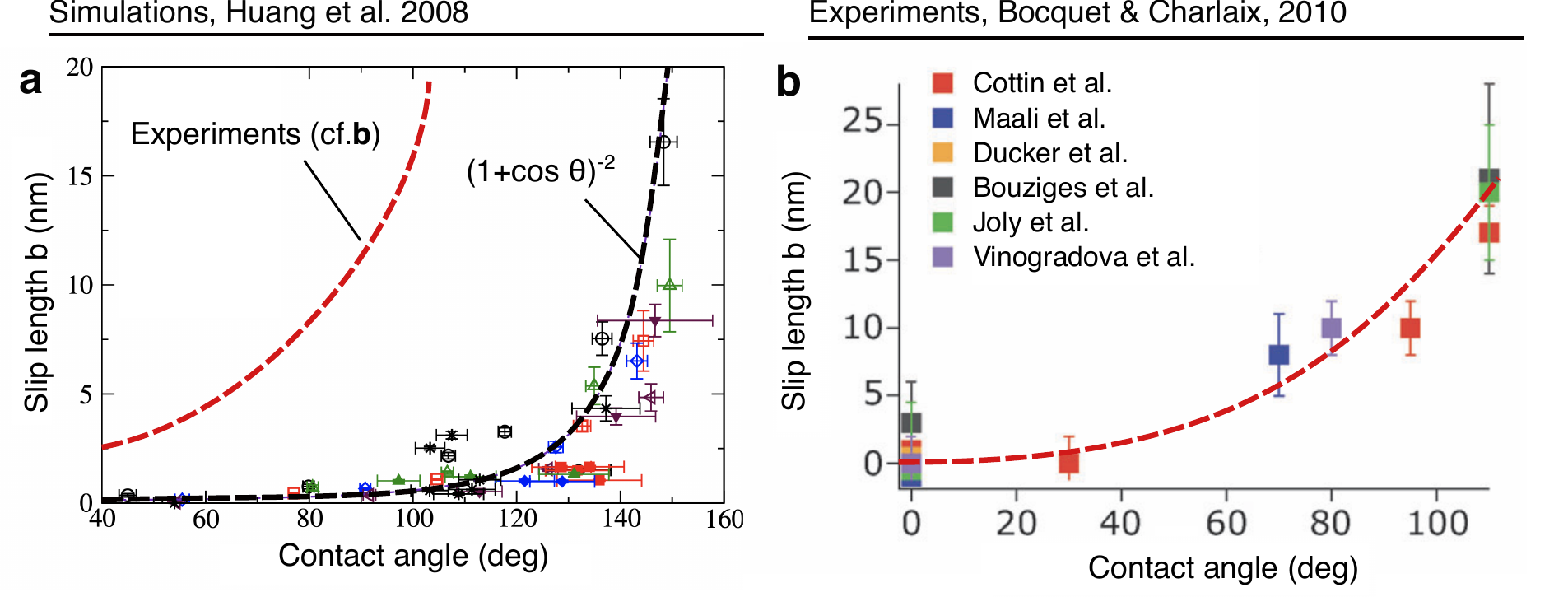} 
		\caption{\textbf{Liquid slippage \textit{versus} contact angle} On panel \textbf{a} reproduced from \cite{huang_water_2008}, we show the results of molecular dynamic simulations on various surfaces and liquid models. The dashed line denotes the following formula : $(1+\textrm{cos}$ $\theta)^{-2}$. On panel \textbf{b}, we reproduce a figure from \cite{bocquet_nanofluidics_2010} showing that current experimental results are in semi-quantitative agreement with the theoretical prediction.}
		\label{Figure_contact_angle}
	\end{figure*}
	
	\paragraph*{Contact angle dependency}
	Across the long story of liquid slippage investigations, the role of hydrophobicity as a slippage enhancer has been often discussed and remains one of the most established findings in the field. On Figure \ref{Figure_contact_angle} \textbf{a} -- reproduced from Ref \cite{huang_water_2008} -- we show the slip length obtained from molecular dynamics simulations on various substrates as a function of the contact angle $\theta$. Considering Equation \ref{eqn:Green_Kubo}, one uses the Young equation to relate the liquid-solid interaction potential to the contact angle and finds $V_{FS}\sim1+cos(\theta)$. Indeed, on panel \textbf{a}, the dashed line scales like $(1+ cos(\theta))^{-2}$ and denotes a 'quasi-universal relationship' between contact angle and slip length, in good agreement with the theory. On panel \textbf{b}, reproduced from Ref.\cite{bocquet2010nanofluidics} we plot selected experimental data clearly showing the strong increase of slippage with hydrophobicity in semi-quantitative agreement with the theoretical scaling. Interestingly, we observe on panel \textbf{a} a systematically higher slippage in experiments than in molecular dynamics simulations.  Despite the success of this theory, it is apparently defied by radius-dependent slippage in carbon nanotubes (\textit{cf.} subsection \ref{subsec:transport}) for which an alternative mechanism has recently been proposed.

	\subsection{Wall fluctuations : Van der Walls friction}\label{subsec:carbon_friction}
		\begin{figure*}[h!]
	\centering
	\includegraphics[width=\linewidth]{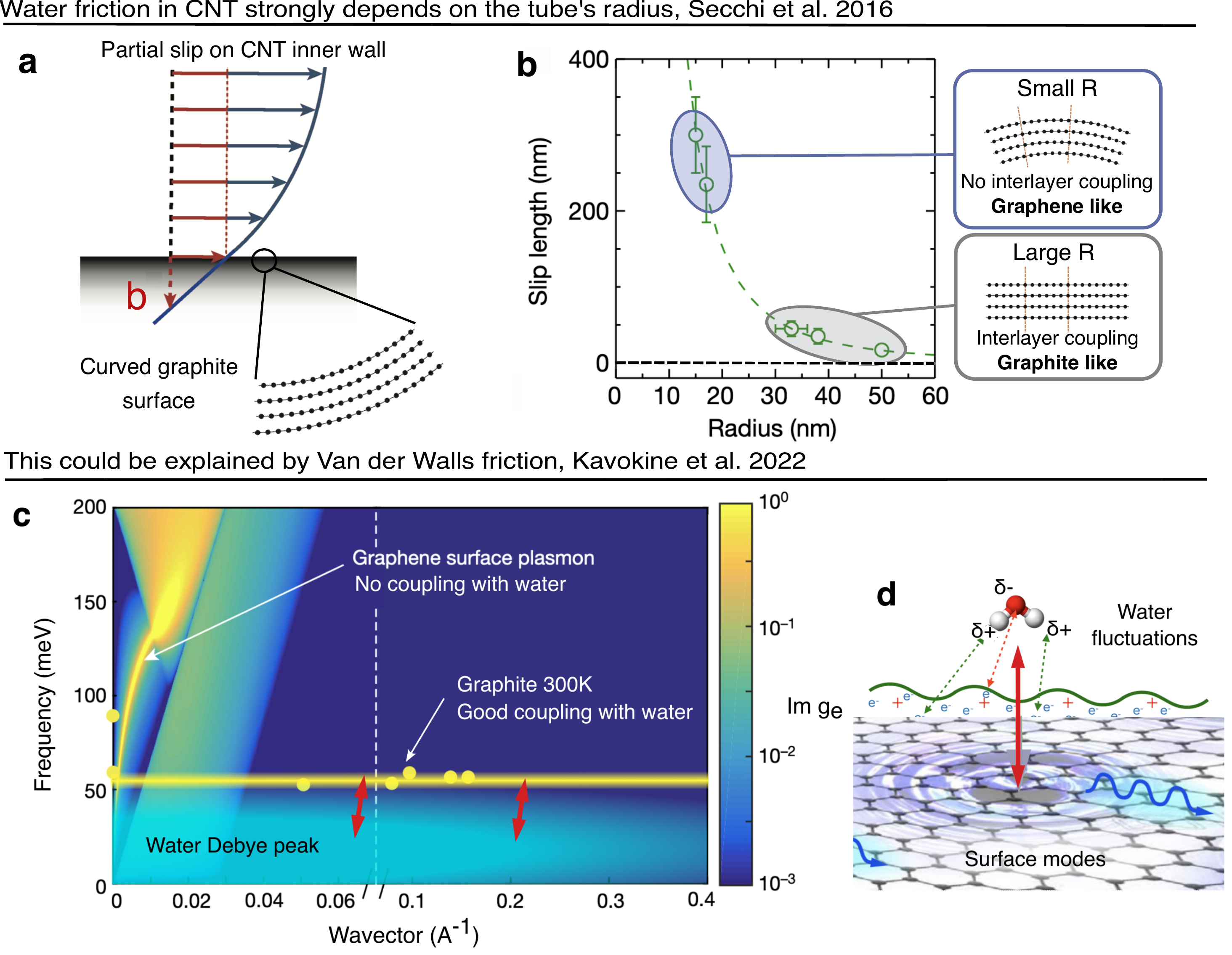} 
	\caption{\textbf{Van der Walls friction} We sketch on panel \textbf{a} liquid slippage on the inner wall of multiwall carbon nanotubes. \textbf{b} The experimental measurements of Secchi \textit{et al.} yield a strong decrease in friction (increase in slippage) for the thinner tubes. This observation is explained by the quantum friction theory of Kavokine \textit{et al.} \cite{kavokine_fluctuation-induced_2022}, using that wider tubes are electronically similar to graphite while narrow tubes, lacking interlayer couplings, are closer to graphene from the electronic point of view. \textbf{c} Electronic excitations for graphite and graphene in the ($q,\omega$) space. Water's Debye peak is added in light blue. The red arrows denote momentum exchange following Eq.\ref{equation:quantum_friction}. \textbf{d} Sketch of the Van der Walls friction which couples water charge fluctuation modes to their solid counterpart through Coulomb interaction. NB : the arrow shown here denotes a surface plasmon but the one of graphite is rather an interlayer, transverse excitation.}
	\label{quantum_friction}
\end{figure*}

	\paragraph*{Multiwall carbon nanotube slippage} The strongly radius-dependent slippage of carbon nanotubes measured by Secchi \textit{et al.} \cite{secchi_massive_2016} was immediatly identified as an extremely puzzling observation: indeed the crystalline structure of the tubes is independent of their diameter to the exception of their curvature, which was shown to have negligible effect on slippage for radii larger than 10 nm \cite{falk2010molecular}. Recently, an original mechanism involving the \textit{electronic} structure of such multiwall carbon nanotubes was proposed by Kavokine \textit{et al.} \cite{kavokine_fluctuation-induced_2022}.
	
	On Figure \ref{quantum_friction} \textbf{a}, we sketch water slippage on the inner surface of a multiwall cabon nanotube, a curved graphite-like interface. Interestingly, it is known that in these tubes, the interlayer electronic couplings strongly depends on curvature. As sketched on panel \textbf{b}, we expect that small radius tubes should behave like a single graphene sheet whereas large radius tubes in which layers are transversally coupled should exhibit the properties of graphite. 
	
	\paragraph*{Theoretical result}  The idea, sketched on Figure \ref{quantum_friction} \textbf{a-d}, is that momentum transfer from liquid flow to the solid wall can be mediated by the Coulomb interaction in addition to the standard molecular collisions. In that scenario, charge fluctuations in the liquid induce a dielectric response in the solid, which in turn acts on the liquid. Under a flow, this reactive force is retarded and tends to oppose movement, effectively dissipating the mechanical energy into the solid's electronic degrees of freedom. For this process to efficiently exchange momentum, a matching in energy and wavevector of dielectric fluctuation modes between the liquid and solid is needed. 	This mechanism is perfectly analogous to the electronic friction between two metallic surfaces which was shown to vanish at the superconducting transition \cite{kisiel2011suppression}.  The Van der Walls friction formula, derived by Kavokine \textit{et al.} in Ref.\cite{kavokine_fluctuation-induced_2022} writes as an integral over the ($q,\omega$) plane of the product of the dielectric surface response functions of respectively the liquid ($g_{\rm w}$) and the metallic electrons ($g_{\rm e}$) multiplied by a $q^3$ contribution: 
	\begin{equation}\label{equation:quantum_friction}
		\lambda_{\rm vdW} = \frac{1}{8\pi^2}\int_{0}^{+\infty}dq \int_{0}^{+\infty} \frac{\hbar d\omega}{k_{\rm B}T}\frac{\hbar q^3}{\textrm{sinh}^2(\frac{\hbar \omega}{2k_{\rm B}T})}\times\frac{g_{\rm e}(q,\omega)g_{\rm w}(q,\omega)}{\vert1-g_{\rm e}(q,\omega)g_{\rm w}(q,\omega)\vert^2}
	\end{equation}
	We mention here that the surface response functions $g_{\rm w}$ and $g_{\rm e}$ are anolog to electromagnetic reflection coefficients and are related to bulk dielectric response functions $\epsilon_{\rm w}$ and $\epsilon_{\rm e}$ by simple analytical formulae.

\paragraph*{Van der Walls contribution to water-graphite friction}
Following the aforementioned argument, we now compare the excitation spectra of graphene and graphite in the frequency-wavector plane ($q,\omega$) (\textit{cf.} panel \textbf{c}). We also add -- in light blue -- water's Debye peak, described in the previous section, and mention that efficient momentum transfer gives a strongly dominant role to high wavevector modes. This matching in high $q$ regions, shown on panel \textbf{c} between water's and graphite's excitation spectra is absent in the case of graphene. We therefore expect a much higher Van-der-Walls friction on graphite than on graphene and explain the large slippage observed in small radius carbon nanotubes. The Van der Walls friction is an additional contribution to the classical roughness-induced friction and shall thus be measurable only on molecularly smooth surfaces. More precisely, the total liquid-friction coefficient writes as the sum of liquid friction on a static wall, free of any fluctuations, and of the Van der Walls term: \begin{equation}\lambda = \lambda_{\rm stat} + \lambda_{\rm vdW}\end{equation} A detailed derivation of the static wall contribution to friction ($\lambda_{\rm stat}$) was estimated from Green-Kubo formula at equation \ref{eqn:Green_Kubo}. Considering the overlap in modes of water and graphite, we compute the Van der Walls formula and find \begin{equation}
	\lambda_{\rm vdW}\sim 1.3 \times 10^{5}\! \textrm{ Pa.s}   \qquad b_{\rm vdW} = \eta/\lambda_{\rm vdW} = 10 \! \textrm{ nm} 
\end{equation} in excellent agreement with experimental findings for water on graphite \cite{maali_measurement_2008}. This calculation suggests that classical friction (static wall) is negligible on graphite. In consequence, the vanishing of electronic friction for graphene should yield a very high slip length.

	\paragraph*{Negative Van der Walls friction ?}
	Recent theoretical works have predicted that Van der Walls friction could become negative, if one assumes an efficient momentum transfer from the liquid to phonons and from phonons to electrons. In this picture where electrons eventally acquire a mean velocity superior to that of the liquid, they drag it in the direction of the flow by the Van-der Wals friction mechanism, resulting in an enhanced slip length  \cite{coquinot_quantum_2023}. This negative Van der Walls friction is a momentum feedback from the solid, mediated by the Coulomb interaction. Although it is, as of now, lacking an experimental confirmation, it could motivate more efforts to engineer low friction metals.
	
	\section{Perspectives in liquid-solid friction}
	In this last section, we dwell on a handful of selected topics which seem to us particularly interesting, dynamic and worthy of investigations in the field of liquid friction.

	\subsection{Super-lubricity or the puzzle of graphene friction}
	The measurement of ultra-fast water flow in carbon nanotubes has sparked a revolution in nanofluidics and given the water-carbon interface a place of its own. The massive slippage suggested by these early measurements is in apparent contradiction with the moderate hydrophilicity of graphite. As detailed above, friction on the inner wall of carbon nanotubes is extremely small and strongly radius dependent. For water on graphene, although some results obtained with an original micro-channel system by Duan \textit{et al.}, suggest a possibly huge slip length of hundreds of nanometers \cite{xie2018fast}, it has still never been measured consistently. Ma \textit{et al.} \cite{li_translucency_2022} (\textit{cf.} Figure \ref{Figure_AFM} \textbf{d}) have demonstrated a slippage transluency of graphene, analog to the previously reported wetting transluency \cite{ondarccuhu2016wettability}. Thus, graphene on a substrate is radically different from its suspended counterpart, from both the contact angle and the slippage perspective. It is thus unsurprising that the inner-wall of carbon nanotubes behaves so strikingly differently to graphene crystals deposited on flat surfaces. To this day, liquid superlubricity remains an exclusive property of carbon nanotubes inner walls and it is a huge challenge to extend it to other surfaces. For technological purposes, it appears necessary to leverage these results in order to fabricate affordable highly permeable CNT-based membranes.
	
	\subsection{Charged interfaces and electrolytes}
	It is crucial for any practical application to consider the effect of surface charge on liquid friction. Real surfaces are often charged in solution and an ionic double layer of the scale of the Debye length is usually present. The nature and mobility of surface charges is often mysterious and concentration dependent.
	
	\paragraph*{Experiments} On the experimental side, very few experiments have linked surface charge and slippage. Streaming measurement (ionic current under pressure drop) measurements in angstrom scale two-dimensional channels have allowed to extract surface charge and infer slippage indirectly \cite{mouterde2019molecular}. In 2-nm carbon nanotubes, a negative surface charge along with a large slip length result in a subtle interplay between pressure driven flows and ionic conductance \cite{marcotte2020mechanically}. Measurements on single carbon nanotubes by Seechi \textit{et al.} have allowed to estimate surface charge as functions of salt concentration \cite{secchi2016scaling} and demonstrated surface charge regulation effects. Unfortunately, the slippage measurements were not extended to salty solutions. The consensus is that graphite is much more slippery than boron-nitride while both have a negative surface charge in agreement with the OH$^{-}$ adsorption obtained in DFT calculations \cite{grosjean2019versatile}. These results are also consistent with the prediction of a much higher mobility of the surface charge on graphite compared to BN. 
	
	\paragraph*{Theory} In a recent numerical study \cite{xie2020liquid}, the authors tried to explain the slippage difference between graphene and BN with a simple model. They compared two types of surfaces with either an homogenous (delocalized) or heterogenous surface charge corresponding respectively to metallic and insulating surfaces. The simulations show that the heterogeneity of surface charge is very detrimental to slippage as it allows ion adsorption and thus slows down the liquid flow close to the surface. Finally, the coupling of ion motion with free electrons in solids through the Van der Walls friction mechanism remains unexplored to this day.
	
	\paragraph*{Perspective} Despite encouraging results, the problem of liquid friction on charged surfaces remains unsolved. Firstly, the nature of the surface charge -- species and type of bound -- is a difficult problem in itself. Then comes its unknown dependency on salt concentration and pH through complex regulation mechanisms which exhibit intriguingly long timescales \cite{robin2023long}. Finally, one has to estimate the friction between the liquid and the surface charge which largely depends on the nature of the material (\textit{cf.} carbon \textit{vs} BN \cite{grosjean2019versatile}). A last problem which will hinder the experimental studies is the possible streaming currents induced by liquid flows on charged surfaces which may lead to complicated flow structures including swirls \cite{hu2012ion}. Concentration gradients and osmotic currents can also couple to mass flow and generate unexpected counter-flow structures which were recently measured with the d-SFA \cite{garcia2016etude}. These multiple couplings between electrostatics, quantum chemistry, and hydrodynamics make it a considerable challenge to study liquid friction on charged surfaces. We believe that experimental efforts in this field are greatly needed. Hydrated friction measurements \textit{versus} salt concentration by Espinoza-Marzal \textit{et al.} \cite{diao2019slippery} are already a promising attempt at systematic investigations.

	\subsection{Liquid slippage and confinement}
	In nanometric channels, hydrodynamic and molecular length scales meet. As deviations from standard hydrodynamics start to develop, it is natural to expect that the friction coefficient could be modified compared to the standard half-plane geometry of liquid slippage.
	
	\paragraph*{Experiments}
	Probing liquid friction at fixed confinement discards all scanning probe techniques which involve fitting a z-spectroscopy curve with a hydrodynamic model. This leaves us with transport experiments only to probe slippage in confined systems. Direct mass flow transport for liquids in nanometric (below 10 nm) confinement and under well-controlled pressure difference has, to our knowledge, never been achieved. Several techniques have been developed to circumvent this constraint in carbon nanotubes as discussed on Figure \ref{figure_CNT}.
	
	\paragraph*{Theory}
	On the theoretical side, the mode coupling approach allows to tackle the problem of confined friction. In a recent study, the authors of Ref.\cite{coquinot2023collective} have computed the friction coefficient for two-dimensional slits as a function of the slit's thickness. Confinement induces a coupling of plasmonic modes in the face to face metallic surfaces yielding a different dispersion to the single surface mode. The effect of this cross-plasmon on friction is relatively weak but not the only consequence of confinement. Indeed, the restricted geometry of a 7 angstrom slit also modifies the dynamical structure factor of water with respect to the bulk. This results in a modified Van der Walls friction force which can result in a 2-fold enhancement of friction coefficient for 7 angstrom slits compared to thicker ones. Note that this mode coupling theory computes only the 'Van der Walls friction' contribution caused by the coupling of water dipoles to free electrons in metallic walls.
	
	\paragraph*{Perspective}
	In parallel to this direction, it is now possible to estimate slippage using an analysis of ion transport measurements under concentration and voltage drops (or pressure and voltage) \cite{emmerich_enhanced_2022,mouterde2019molecular}. Eliminating the need for exquisite mass flow measurements, this technique trades experimental difficulty for a challenging rationalization and several hypotheses. This approach is particularly promising for studies of slippage in ultra-confined systems and has been applied to that of small-diameter carbon nanotubes \cite{cui2023enhanced}, angstrom-scale two-dimensional channels \cite{mouterde2019molecular}, and activated carbon nano-conduits \cite{emmerich_enhanced_2022}. Despite these interesting proposals, true friction measurements for confinements below 10 nm are still lacking. We point out that, more generally, high resolution mass flow measurement techniques in ultimate confinement are extremely rare but would make a huge contribution not only to the confined-friction problem but to nanofluidics in general.
	
	\begin{figure*}[h!]
	\centering
	\includegraphics[width=\linewidth]{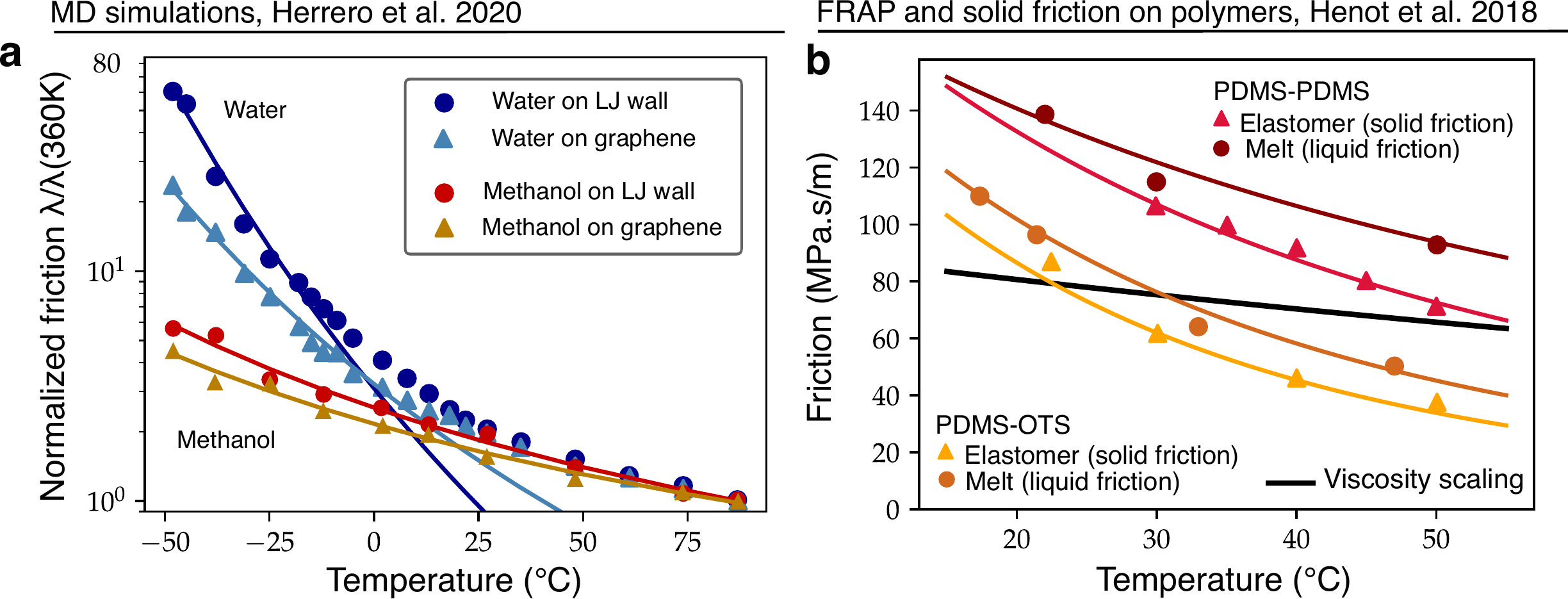} 
	\caption{\textbf{Supercooled liquid-solid friction} On panel \textbf{a}, we replot the data from \cite{herrero_fast_2020} where the authors use molecular dynamics simulations to compute the friction coefficient and slip length of water and methanol on graphene and a Lennard-Jones wall \textit{versus} temperature. They observe a strong increase of friction upon cooling, the solid lines are fits with an Arrhenius law. Note that the friction coefficient is normalized by its value at 360K. On panel \textbf{b}, we display the experimental results from \cite{henot_temperature-controlled_2018}. In this work, the authors use a FRAP setup to measure the dependency of friction coefficient on temperature for polymer melts of various molar masses.}
	\label{supercooled_liquids}
\end{figure*}
	
	\subsection{Liquid dynamics and non-Newtonian interfaces}
	
	As evidenced in Green-Kubo equation Eq. \ref{eqn:Green_Kubo}, liquid friction intrinsically depends on molecular timescales needed for force correlations to relax. Supercooled liquids offer a unique platform to tune molecular timescales on demand. Indeed, a mere temperature drop allows to drastically shift the liquid's dielectric relaxation peak or in other words its molecular relaxation time which is close to the Maxwell time $\tau \sim \eta/G_{\infty}$ ($G_{\infty}$ is the high-frequency elastic modulus). These measurements would be especially interesting in the Van der Walls friction framework, in which an overlap of liquid and solid excitations in the ($q,\omega$) space enhances friction. Moreover, like many biological fluids, ionic liquids are often already supercooled at room temperature and are routinely confined in battery electrodes, super-capacitors \cite{chmiola2006anomalous} and membranes used to harvest temperature gradients \cite{pascual2023waste}.
	
	\paragraph*{Interfacial visco-elasticity ?}
	Finally, in slow liquids under shear, the Newtonian hypothesis may no longer hold. Whereas the complex and frequency dependent rheology of the bulk is often well known (complex frequency dependent viscosity $\tilde{\eta}(\omega)$), it is a fully open question to know how the interface behaves at high frequency. First experiments in poly-electrolyte solutions have shown that liquid-solid friction remained real \cite{cross_wall_2018} whereas the bulk deviated strongly from newtonian behavior but we could assume that the interfacial friction coefficient becomes complex and frequency dependent $\tilde{\lambda}(\omega)$ at high shear rate. 
	
	\paragraph*{First results}
	On Figure \ref{supercooled_liquids}, we show the results of two studies of supercooled liquid friction on solid surfaces. On panel \textbf{a}, we plot the temperature-dependent friction coefficient  extracted from molecular dynamics simulations of water and methanol on either graphene or a Lennard-Jones wall \cite{herrero_fast_2020}. The solid lines are fits with an Arrhenius law. On panel \textbf{b}, we reproduce experimental results from \cite{henot_temperature-controlled_2018} on polymer melts. In this work, the authors report an Arrhenius dependency for both viscosity and friction. Having in mind that viscosity $\eta$ describes liquid-liquid friction while $\lambda$ denotes the interfacial friction, these two different activation energies account for two separate mechanisms for friction in these two situations. It is interesting to note that a variations of the slip length $b=\eta/\lambda$ implies a bulk-surface decoupling upon cooling. In this latter study, the authors compared for each surface the polymer melt's liquid friction $k_{melt}=\eta/b$ with the solid friction $k_{elastomer}=\sigma_\parallel/V_\parallel$ of the corresponding reticulated polymer. When both are plotted together as a function of temperature on Figure \ref{supercooled_liquids} \textbf{b}, it is striking that they have roughly the same temperature dependency and order of magnitude. This observation strengthens the claim that liquid friction depends on the monomers only and that its dynamical behavior is radically decoupled from the bulk's for which viscosity scales as the chain's length to a power close to 3.5 \cite{colby1987melt}. Finally, we mention a recent investigation, by the authors, of the friction of supercooled glycerol on mica. Using a dynamic CP-AFM, we reported drastic deviations to Arrhenius behavior and proposed that the strong shift of the liquid's collective modes upon cooling is tuning liquid-solid friction in a similar way to the Van der Walls mechanism described above \cite{lizee2024anomalous}.


	
	\section{Conclusion}

		Liquid-solid friction is now established as a mature field of research with enormous technological relevance and several key experimental techniques. Due to its extreme sensitivity to surface roughness, slippage is rare for water. Nevertheless, a handful of materials exhibit drastically reduced friction, which suggests that highly permeable -- yet selective -- membranes may be available in the near future. The emerging Van der Walls friction for liquids promises to control slippage by tuning electronic properties in the solid state. We argue that the respective influence of surface charge, nanometric confinement and liquid supercooling are open questions that deserve special attention and for which the experimental techniques are mature.
		
		Finally, we emphasize that, like the combined AFM measurements of liquid and hydrated solid friction, which provide the activation energy $E_{\rm a}$ and the slip length $b$, the study of polymer friction for both the melt (liquid) and the reticulated polymer (solid) has provided groundbreaking physical insights. We argue that the coupling of liquid and solid friction -- which now appear to be complementary -- will be key to unravelling the dynamical processes at the liquid-solid interface.

\bibliography{Friction_chapter_figs/slip_chapter.bib}
\end{document}